\documentclass{iopjournal}
\usepackage{multirow}
\usepackage{boldline,multirow}
\usepackage{cite}
\usepackage{amsmath,amssymb,amsfonts}
\usepackage{algorithmic}
\usepackage{algorithm}
\usepackage{graphicx}
\usepackage{textcomp}
\usepackage{xcolor}
\usepackage{booktabs}
\usepackage{amsmath}
\usepackage{array} 
\usepackage{subcaption}
\usepackage{bm}
\usepackage{breakurl}
\usepackage{mathrsfs}   
\usepackage{calligra}   
\usepackage{url}
\usepackage{hyperref}
\usepackage{threeparttable}
\usepackage{makecell}
\usepackage[normalem]{ulem}
\usepackage{soul}
\useunder{\uline}{\ul}{}

\newcolumntype{C}[1]{>{\centering\arraybackslash}p{#1}}
\def\BibTeX{{\rm B\kern-.05em{\sc i\kern-.025em b}\kern-.08em
    T\kern-.1667em\lower.7ex\hbox{E}\kern-.125emX}}

\begin{document}

\articletype{Paper} 

\fancyhead[R]{\@Nguyen {\it et al}}

\title{Flexible Genetic Algorithm for Quantum Support Vector Machines}

\author{Nguyen Minh Duc$^{1,3}$\orcid{0000-0000-0000-0000}, Vu Tuan Hai$^{2,3,*}$\orcid{0000-0000-0000-0000}, and Le Bin Ho$^{4,5,^\dagger}$\orcid{0000-0002-8816-4450} and Lan Nguyen Tran$^{1,3,\dagger\dagger}$\orcid{0000-0000-0000-0000}}

\affil{$^1$University of Science, Vietnam National University, Ho Chi Minh City 700000, Vietnam.}

\affil{$^2$University of Information Technology, Vietnam National University, Ho Chi Minh City 700000, Vietnam.}

\affil{$^3$Vietnam National University, Ho Chi Minh City 700000, Vietnam.}

\affil{$^4$Frontier Research Institute for Interdisciplinary Sciences, 
Tohoku University, Sendai 980-8578, Japan.}

\affil{$^5$Department of Applied Physics, 
Graduate School of Engineering, 
Tohoku University, 
Sendai 980-8579, Japan.}

\email{$^*$haivt@uit.edu.vn, $^\dagger$binho@fris.tohoku.ac.jp, $^{\dagger\dagger}$tnlan@hcmus.edu.vn}


\begin{abstract}
Quantum Support Vector Machines (QSVM) is one of the most promising frameworks in quantum machine learning, yet their performance depends on the design of the feature map. Conventional approaches rely on fixed quantum circuits, which often fail to generalize across datasets. To address this limitation, we propose GA-QSVM, a hybrid framework that employs Genetic Algorithms (GA) to automatically optimize feature maps. The proposed method introduces a configurable framework that flexibly defines the evolutionary parameters, enabling the construction of adaptive circuits. Experimental evaluation of datasets, including Digits, Fashion, Wine, and Breast Cancer, demonstrates that GA-QSVMs achieve a comparable accuracy compared to classical SVMs and standard QSVMs. Furthermore, transfer learning results indicate that GA-QSVM's circuits generalize effectively across datasets. These findings highlight the potential of evolutionary strategies to automate and enhance kernel design for future quantum machine learning applications.
\end{abstract}

\section{Introduction}

Quantum Machine Learning (QML) \cite{PhysRevLett.122.040504, Schuld03042015} has emerged as a promising field with the potential to offer computational advantages over classical methods. Among various QML algorithms, the Quantum Support Vector Machine (QSVM) \cite{PhysRevLett.113.130503, Gentinetta2024complexityofquantum} is particularly notable, as it translates a well-understood classical Support Vector Machine (SVM) into the quantum domain. For the dual problem, a classical SVM has a complexity of $\mathcal{O}(M^2)$ for an $M$-size dataset. In contrast, the quantum dual problem requires $\mathcal{O}\!\left(\frac{M^{4.67}}{\epsilon^2}\right)$ quantum circuit evaluations to achieve an $\epsilon$-accurate solution \cite{Gentinetta2024complexityofquantum}. This indicates that the quantum dual approach grows more rapidly with problem size than the classical method.
The power of a QSVM lies within its quantum kernel $K$ between two quantum states $|\psi(x_i)\rangle$ and $|\psi(x_j)\rangle$, which measures the similarity between them. These states are prepared by the quantum feature map $\phi$ through a unitary $U_\phi(\bm\theta)$.

The primary challenge in QSVM is the design of the parameterized circuit used in the feature map. Complex circuits can capture data patterns, but they face two significant issues. First, in high-dimensional feature spaces, the inner product between data vectors $x_i^{\intercal}x_j$ concentrates around a constant $\epsilon$, which reduces the variability of the kernel matrix, making it nearly uniform and thus difficult to train. Second, overly expressive circuits often lead to poor generalization, a phenomenon analogous to overfitting in classical machine learning \cite{sppss2019, Huang2021}. Currently, QSVM parameterized circuits often rely on fixed ansatzes, such as ZZFeatureMap \cite{PhysRevA.101.032308}, Hardware Efficient Embedding (HEE) \cite{Thanasilp2024, Leng2025expandinghardware} or Instantaneous Quantum Polynomial (IQP)-style circuits \cite{recio2025iqpopt, PRXQuantum.6.020338}, which may not be optimal for a general purpose. 

To address the challenge of manual circuit design, previous works used Genetic Algorithm (GA) to automatically discover effective circuits \cite{creevey2023kernel, Creevey2023_GASP, PRXQuantum.5.040339,linh2025advancingquantumprocesstomography, ALTARESLOPEZ2024122984, Yan2024, Senokosov_2024}. GA is a powerful metaheuristic for searching in many quantum problems, such as state preparation \cite{Creevey2023_GASP, PRXQuantum.5.040339}, tomography \cite{linh2025advancingquantumprocesstomography}, image processing \cite{ALTARESLOPEZ2024122984, Yan2024, Senokosov_2024}, and quantum compilation \cite{Hai_2024}. In particular, in the QSVM framework, there have been several works employing GA to search for optimal feature maps. In Ref.~\cite{creevey2023kernel}, the authors applied GA to search gate-sequence encodings, showing that GA-discovered circuits can outperform hand-designed circuits while revealing a positive relationship between classifier test accuracy and kernel entropy. Pello {\it et al.} \cite{Pellow-Jarman2024} demonstrated that the kernel-target alignment is suitable for the direct accuracy evaluation inside an NSGA-II \cite{996017} based GA. Those authors further accelerated the fitness evaluation by using an approximation to kernel-target alignment. Gate parameters are optimized with classical optimizers, producing compact maps with improved alignment. Recently, Wang et al. have studied entanglement explicitly with a multi-objective GA that separates fitness contributions from local and non-local gates ~\cite{Wang2025}. Their work showed that (i) optimal circuits tend to include more non-local gates rather than eliminating them, and (ii) the multi-objective selection yields the Pareto front representing a set of optimal trade-offs where improving accuracy leads to an increase in kernel size.

Existing GA-QSVM approaches still have several limitations. First, their performance often highly depends on the benchmark dataset. As a result, QSVM does not generally outperform classical SVMs on various tasks \cite{creevey2023kernel}. Second, previous GA procedures lacked a normalization mechanism to ensure that evolved solutions remain aligned with the original objective \cite{Pellow-Jarman2024}. Finally, previous methods lacked well-defined evolutionary strategies to improve fitness in the next generation \cite{Wang2025}. To address these limitations, we propose a configurable GA framework that allows the dynamic definition of fitness function, normalizer, generator, and GA operations. Our framework can automatically adjust resources, such as increasing the circuit depth or adding more gates, when GA fails to find an optimal solution. For evaluation, we select four toy datasets that are suitable for simulating QSVM on a classical computer.. We then conduct a comparison between classical SVMs,  QSVMs with standard feature maps, and our proposed GA-QSVMs. Especially, we also assess the performance of our GA-QSVM with different types of kernels, including Fidelity Quantum Kernel (FQK) and Projected Quantum Kernel (PQK) \cite{PhysRevA.109.042612, huangPowerDataQuantum2021}.


The remainder of this paper is organized as follows. Section~\ref{sec:related_work} reviews the related works on QSVM and GA applications in quantum computing. Section~\ref{sec:background} presents the background of GA, classical SVM, and quantum kernels. Section~\ref{sec:proposal} describes the proposed GA-QSVM methodology, including the evolutionary design of quantum circuits and metadata configuration. Section~\ref{sec:evaluation} provides the evaluation results on benchmark datasets, hyperparameter analysis, and comparison with classical and existing quantum SVMs. Finally, Section~\ref{sec:conclusion} concludes the paper and discusses potential future research directions.

\section{Related works} \label{sec:related_work}

Recently, QSVM has emerged as one of the prominent QML approaches. Several works \cite{rebentrostQuantumSupportVector2014, havlicekSupervisedLearningQuantum2019} provided the foundation for applying quantum-enhanced feature spaces to supervised learning tasks. These early studies demonstrated how quantum feature maps could transform classical data into high-dimensional Hilbert spaces, enabling the classification of non-linear patterns that are difficult to capture with classical kernels. Later works have shown that the QSVM performance depends strongly on the dataset \cite{babuEntanglementenabledQuantumKernels2025, huangPowerDataQuantum2021}. While QSVM and classical SVM perform comparably for simple datasets, QSVM has been found to outperform the classical one for complex datasets thanks to the better representational of quantum feature maps in these cases. Beyond benchmarking, the QSVM framework has been successfully applied to real-world problems. For example, QSVM can solve the vehicle routing problem \cite{mohantySolvingVehicleRouting2024}, air pollution prediction \cite{farooqEnhancedApproachPredicting2024}, DNA sequence similarities \cite{shiCompareSimilaritiesDNA2025}, and handling data with underlying group structures through covariant kernel formulations \cite{glickCovariantQuantumKernels2024}.

Despite the above-mentioned advances, some major challenges remain. Mapping classical data into quantum states has proven to be non-trivial, with subtleties that affect both trainability and performance \cite{thanasilpSubtletiesTrainabilityQuantum2023}. As the same in classical SVM, overfitting is another recurring issue in QSVM \cite{petersGeneralizationOverfittingQuantum2023, jerbiQuantumMachineLearning2023a}. Furthermore, a key challenge in QSVMs lies in the kernel design. Trainable quantum kernels have been proposed to adapt the kernel based on datasets, improving the ability of overall generalization \cite{hubregtsenTrainingQuantumEmbedding2022, xuQuantumClassifiersTrainable2025}. Complementing this line of work, Suzuki et al. \cite{suzukiAnalysisSynthesisFeature2020} analyzed the expressivity of quantum feature maps, providing insights into how different kernels affect the performance of classifiers. Gil-Fuster \textit{et al} showed that one can improve the performance of QSVM by designing new quantum kernels\cite{gil-fusterExpressivityEmbeddingQuantum2024}. Recently, the GA-QSVM approach introduced in Ref.~\cite{creeveyKernelAlignmentQuantum2023} leveraged GA to optimize kernel structure, representing a step toward automating the kernel selection problem. This hybrid evolutionary-quantum strategy enhances the accuracy of QSVMs, particularly when dealing with complex datasets.



\section{Background} \label{sec:background}

\subsection{Genetic algorithms}

GAs are classes of evolutionary optimization techniques inspired by the biological process of natural selection. They operate by maintaining a population that evolves over generations according to a defined fitness function. A population consists of multiple candidates, forming the search space for the algorithm. The structure of each candidate is encoded as a chromosome serving as its genetic representation. Within each chromosome, a gene corresponds to a specific component of the overall solution \cite{ACAMPORA2023110296, Alhijawi2024}.

The evolution of the population proceeds by repeating selection, crossover, mutation, and replacement operations. In the selection phase, candidates with higher fitness values are more likely to be chosen as parents for the next generation. The crossover operation then combines parts of two or more parent chromosomes to produce offspring, thereby promoting the exchange of better genes. Mutation introduces small random changes to candidate genes, preserving the genetic diversity and helping the algorithm avoid local minima. Finally, the replacement determines which candidates from the current and new generations will survive to become the next population. Through this iterative process, GA gradually refines candidate solutions, converging toward an optimal or near-optimal result. Their ability to search large, complex, and non-convex spaces without relying on gradient makes them particularly effective in optimization \cite{Dong2021, Wang2023, Zhang2025}.

Conventional GAs used in existing GA-QSVM frameworks \cite{creevey2023kernel, Pellow-Jarman2024, Wang2025} often face limitations due to their static evolutionary mechanisms, leading to a large deviation from the targeted optimization objective. Earlier GA-QSVM implementations also typically lacked a normalization mechanism to ensure that evolved individuals remain consistent with the targeted fitness landscape, resulting in unstable evolution across generations. Additionally, GA frameworks do not have adaptive strategies to enhance fitness progression effectively in generations. To overcome these issues, recent studies have introduced Adaptive Genetic Algorithms \cite{Gen2023, LI2024101529} (AGAs) to dynamically adjust their parameters, such as resources, mutation rate, crossover probability, and selection pressure, based on the current population state. These adaptive mechanisms enhance convergence robustness across various environments.












\subsection{Classical Support Vector Machine}

SVM\cite{708428, Noble2006} is a popular and effective supervised approach to linear classification. The prediction is made using the function of the inner product of $x$ and a weight vector $w \in \mathcal{X}:f^*(x) = \text{sign}(f(x) + b)$, where $f(x) = w \cdot x$, defining a separate hyperplane with orientation and offset controlled by $w$ and  $b$, respectively.

To handle non-linearly separable data, a soft margin is employed, and a hinge loss is then traded off with the norm of $w$ when training SVM models. We can see it as a regularization part to avoid overfitting. For complex and nonlinear problems, data points are mapped from the input space $\mathcal{X}$ to a higher-dimensional space $\tilde{\mathcal{X}}$ using a mapping function $\phi:\mathcal{X} \rightarrow \tilde{\mathcal{X}}$. In this new space, a linear decision boundary is found. The training process for a soft-margin SVM is defined by the following primal optimization problem:

\begin{align}
    \min_{w, b, \epsilon} \quad \frac{1}{2} \big\|w\big\|_2^2 + C \sum_{i=1}^n \epsilon_i
\end{align}

subject to:  

\begin{align}
    \epsilon_i \ge 0, \text{ and } y_i(w \cdot \phi(x_i) + b) \ge 1 - \epsilon_i \quad \forall i=1, \dots, n,
\end{align}

where the parameter $C > 0$ determines the trade-off between increasing the margin size and ensuring that the ${x}_{i}$ lie on the correct side of the margin.

The kernel trick is used to avoid explicitly computing the mapping $\phi(x_i)$. By using a kernel function, $K(x_i, x_j)$, we can calculate the inner product of vectors in $\tilde{\mathcal{X}}$ directly from the original input vectors. This allows the problem to be solved using its dual formulation, expressed as:

\begin{align}
     \max_{\beta} \quad L(\beta) = \sum_{i=1}^n \beta_i - \frac{1}{2} \sum_{i=1}^n \sum_{j=1}^n \beta_i \beta_j y_i y_j K(x_i, x_j),
\end{align}

subject to:

\begin{align}
    \sum_{i=1}^n \beta_i y_i = 0, \text{ and } 0 \le \beta_i \le C \quad \forall i=1, \dots, n.
\end{align}

Classical SVMs fundamentally rely on user-chosen kernels whose inductive bias may not capture higher-order, non-local correlations in the data. In practice, performance depends heavily on heuristic kernel selection and bandwidth tuning; poorly tuned kernels either overfit (small bandwidth) or ``wash out'' structure (large bandwidth), and hyperparameters interact in nontrivial ways \cite{https://doi.org/10.1002/qute.202300298}. Moreover, kernel methods scale at least quadratically in sample size due to the $N\times N$ Gram matrix, creating time/memory bottlenecks for large $N$ and complicating cross-validation and model selection. Even sophisticated classical kernels or twin-SVM variants cannot fully avoid these scaling limits \cite{Egginger2024}.


\subsection{Quantum Support Vector Machines}

\begin{algorithm}[t]
\caption{QSVM $\mathcal{F}$}
\begin{algorithmic}[1]
\REQUIRE Dataset $D$, quantum circuit $\mathbf{g}$
\STATE Construct quantum kernel $K$ with $\mathbf{g}$ as feature map
\STATE Construct $\{x_{\text{train}}, y_{\text{train}}, x_{\text{test}}, y_{\text{test}}\}$ from $D$.
\STATE Train the classifier $\mathbf{C}_{K}$ with $\{x_{\text{train}}, y_{\text{train}}\}$
\STATE $y_{\text{pred}} \gets \mathbf{C}_{K}(x_{\text{test}})$
\RETURN Accuracy between $\{y_{\text{test}}, y_{\text{pred}}\}$
\end{algorithmic}
\label{algo:qsvm_algo}
\end{algorithm}

QSVMs are quantum versions of classical SVMs. As defined in Algorithm~\ref{algo:qsvm_algo}, they leverage quantum principles to handle high-dimensional data. QSVMs define kernels via state fidelity in exponentially large Hilbert spaces using data-encoding circuits. Carefully designed circuits can embed classically hard-to-compute similarities, potentially improving generalization for certain tasks \cite{https://doi.org/10.1002/qute.202300298}. 
In QSVM, the datapoint $x_i$ is encoded into a quantum state $|\psi(x_i)\rangle$. This mapping is achieved through a quantum feature map $U(\bm\theta,x)$ applied to an initial state $|\bm 0\rangle$:

\begin{align}
    |\psi(x)\rangle = U(\bm\theta,x)|\bm 0\rangle.
\end{align}


The widely-used QSVM kernel is FQK $K$, defined as the fidelity between two quantum states encoded by the circuit $\mathbf{g}$, denoted as $K^{FQK}_{\mathbf{g}}(x_i, x_j)$:
\begin{align}
    K^{FQK}_{\mathbf{g}}(x_i, x_j) = |\langle\psi(x_i)|\psi(x_j)\rangle|^2.
\end{align}
Each input $x_i$ is then fed into classifier $\mathbf{C}_{K}$, which returns the corresponding label $y_i$.


To overcome the drawback of standard FQK suffering from small geometric differences when the effective dimension $d$ is large, PQK has been recently developed by Huang et. al. \cite{Huang2021}. PQK projects quantum states into approximate classical representations, such as reduced physical observables or classical shadows, enabling better generalization in low-dimensional classical spaces. The simple form of PQK measures the one-particle reduced density matrix (1-RDM) for each qubit: $\rho_k(x_i) = \mathrm{Tr}_{j \neq k}[\rho(x_i)]$ and is defined as

\begin{align}
    K^{PQK}_{\mathbf{g}}(x_i, x_j) = \exp \left(-\gamma \sum_k \|\rho_k(x_i) - \rho_k(x_j)\|_F^2 \right),
\end{align}
which is known as the Frobenius norm between two 1-RDM. PQK can be efficiently computed via classical shadows, providing a practical route for quantum-enhanced learning with fewer measurements. Recent works have shown that quantum kernels are efficient in multiclass classification when quantum circuits and their hyperparameters are chosen properly \cite{PhysRevA.111.062410, Thanasilp2024, Schnabel2025}. In particular, QSVMs can bypass classical SVM limitations when (i) the quantum feature map targets data structures inapplicable to classical kernels, and (ii) hyperparameters are carefully managed \cite{Schnabel2025}. The GA approach with dynamically designing quantum circuits for the kernel is expected to solve the two above-mentioned problems.

\section{Methodology} \label{sec:proposal}

\begin{table}[]
\caption{GA-QSVM terminology mapping}
\centering
\label{tab:ga_qsvm_terminologies_booktabs}
\resizebox{.75\textwidth}{!}{
\begin{tabular}{|p{2cm}|p{5cm}|p{5cm}|}
\hline
\textbf{GA} & \textbf{GA-QSVM} & \textbf{Description} \\ \hline
Gene & Quantum gate $g$ & Unitary operator samples from Pool gates \\
Candidate & Parameterized quantum circuit $\mathbf{g}$ & A set of $g$ \\ 
Population & Generation $G^{(i)}$ & A set of $\mathbf{g}$ \\ 
Fitness & Fitness & Performance metric \\ \hline
\end{tabular}}
\end{table}

Our approach combines GA and QSVM into a hybrid workflow as a two-level optimizer. QSVM serves as a low-level optimizer to find an optimal hyperplane, whereas GA is the high-level optimizer to find the best circuit, which strongly affects the QSVM optimization.

\subsection{GA-QSVM procedure}

\begin{figure}[t]
    \centering
\includegraphics[width=0.75\textwidth]{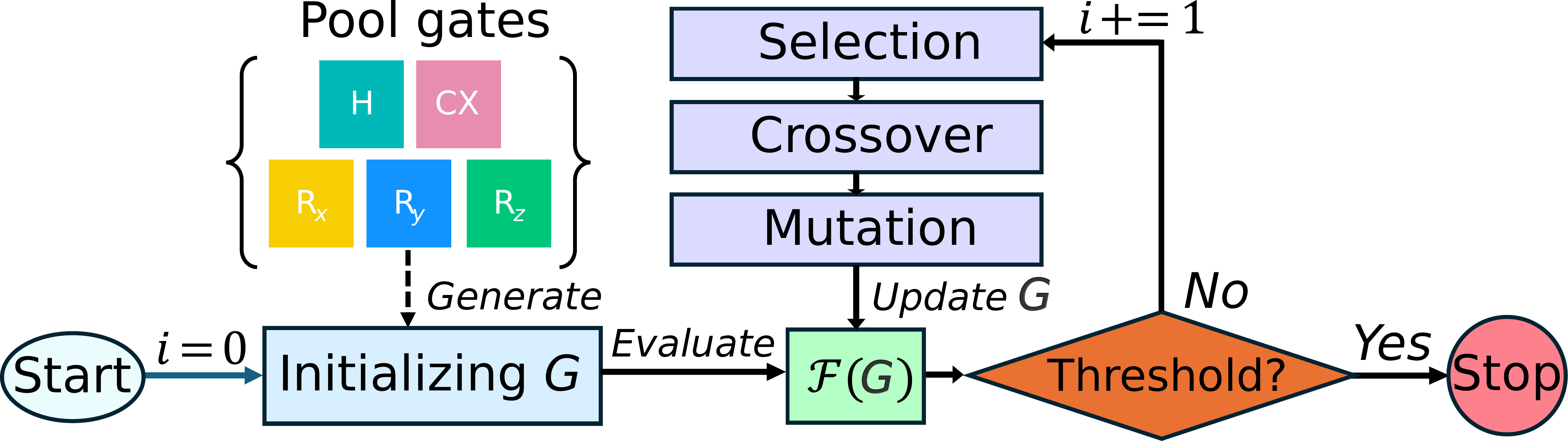}
    \caption{The flowchart of the GA-QSVM procedure. At generation $i$, we evaluate set $G$ then update this set until at least one fitness value meets the threshold.
    }
    \label{fig:system-architecture}  
\end{figure}

\begin{figure*}[t]
    \centering
\includegraphics[width=0.98\textwidth]{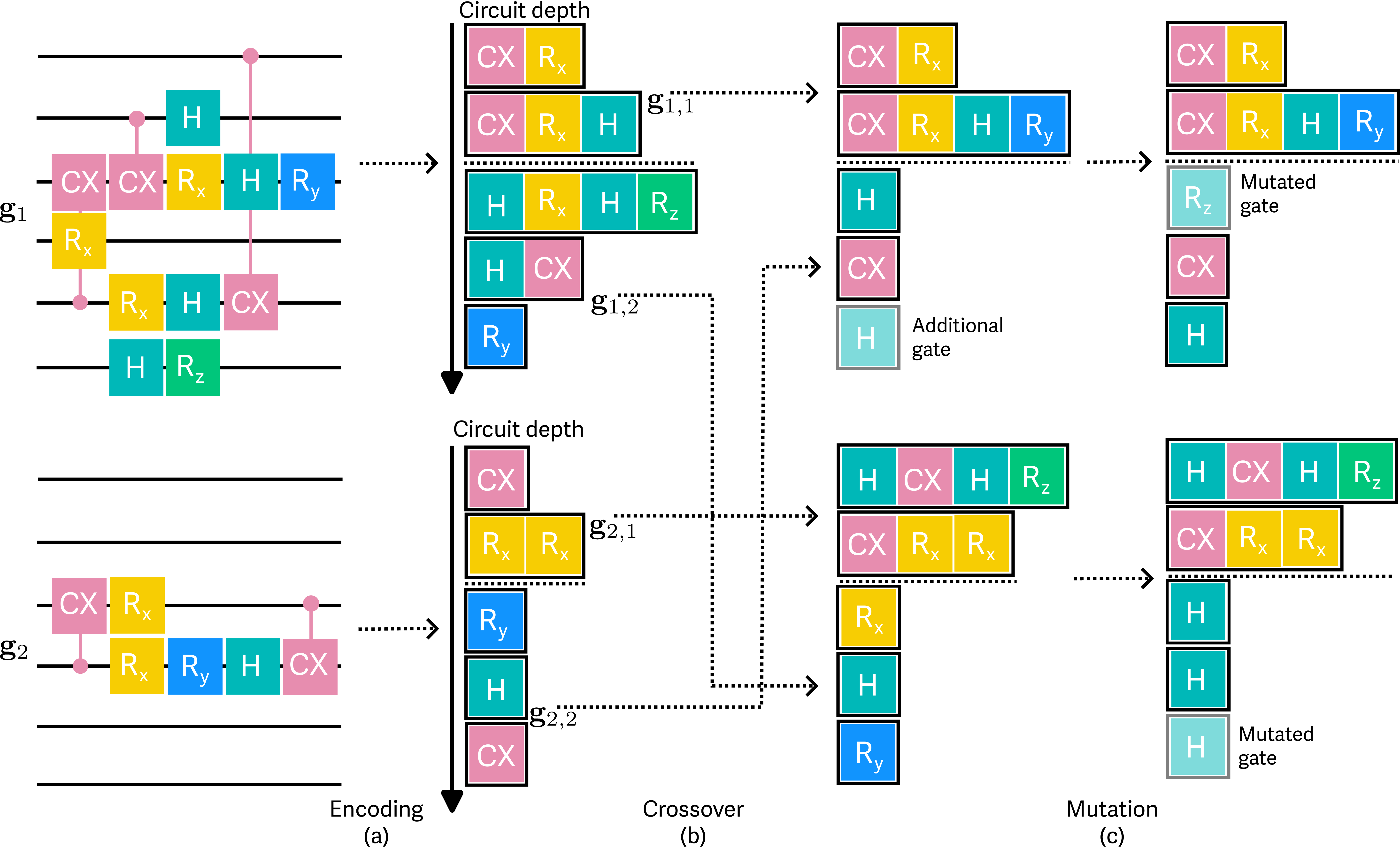}
    \caption{GA operations on quantum circuits. (a) Two 6-qubit quantum circuits $\{\mathbf{g}_1,\mathbf{g}_2\}$ as encoded as genome by circuit depth. (b) Two quantum circuit are crossover: $\mathbf{g}_1$ and $\mathbf{g}_2$ are divided into $\{\mathbf{g}_{1,1},\mathbf{g}_{1,2},\mathbf{g}_{2,1},\mathbf{g}_{2,2}\}$ then $\mathbf{g}_{1,1}$ combines with $\mathbf{g}_{2,2}$, $\mathbf{g}_{2,1}$ combines with $\mathbf{g}_{1,2}$. (c) The offspring then randomly mutate. Note that if the offspring are not followed by a normalized condition, the addition/truncation gate's operation may be applied due to the normalization condition.}
    \label{fig:example}  
\end{figure*}

\begin{algorithm}[t]
\caption{GA for QSVM task}
\begin{algorithmic}[1]
\REQUIRE Fitness function $\mathcal{F}$, metadata $\mathcal{M}$
\STATE $G^{(0)}\gets \text{Generator}(\mathcal{M})$
\FOR{$i$ in $[0\ldots n_{\text{generation}}]$}
\STATE ${F}\gets [\mathcal{F}(\mathbf{g})\;\text{for} \;\mathbf{g}\;\text{in}\;G^{(i)}]$ 
\IF{$\exists f\in F, f<\mathcal{M}\tau$}
    \RETURN $G^{(i)}$
\ENDIF
\STATE $\{G^{(i)}_{\text{father}},G^{(i)}_{\text{mother}}\}\gets \text{Selection}(G^{(i)}, \text{Metadata})$
\STATE $G^{(i+1)}\gets \text{Mutation}(\text{Crossover}(\{G^{(i)}_{\text{father}},G^{(i)}_{\text{mother}}\},\ldots),\ldots)$
\ENDFOR
\RETURN $G^{(i)}$
\end{algorithmic}
\label{algo:fitness}
\end{algorithm}

Our GA-QSVM procedure is presented in Figure.~\ref{fig:system-architecture}. Each candidate in the population represents a circuit composed of a sequence of gates from a pool. Notice that previous works included $R_X$, $R_Z$, and CX gates restricted to 90-degree ($\sqrt{\text{X}}$), but lack of the Y-axis rotation gate \cite{Pellow-Jarman2024}. In this work, to enhance the flexibility, we involve $R_x(\theta), R_y(\theta), R_z(\theta), \text{H}, \text{CX}$ gates.

The corresponding fitness $\mathcal{F}(\ldots)$ of each circuit is determined by its performance for a given dataset $D$ that is separated into $\{D_{\text{train}}, D_{\text{test}}\}$. As given in Algorithm~\ref{algo:fitness}, for each circuit in the population, we perform QSVM using the corresponding quantum kernel. QSVM is trained on $D_{\text{train}}$ and its classification accuracy is measured on $D_{\text{test}}$, this score serves as the fitness:

\begin{align}
\text{Accuracy} = \frac{1}{|y_{\text{test}}|} \sum_{j=1}^{|y_{\text{test}}|} \mathbb{I}(\hat{y}_{\text{test}_j} == y_{\text{test}_j}),
\end{align}

where $\mathbb{I}(\cdot)$ is the indicator function, which is $1$ if the predicted label $\hat{y}$ matches the true label $y$ and $0$ otherwise. 


At the end of each generation, the circuits are ranked by their fitness. The evolution then proceeds with three steps: selection, crossover, and mutation as depicted in Figure.~\ref{fig:example}. We employ the elitist selection strategy, in which the best-performing circuits of the population are chosen to be parents for the next generation. Then, it was divided into the father set and the mother set. The number of eliminated circuits is set to $n_{\text{circuit}}/2$ by default, meaning that the population is kept over generation if the initial population is divisible by $4$.
\begin{align}
    \{G_{\text{father}}^{(i)}, G_{\text{mother}}^{(i)} \}\gets \text{Selection.Elitist}(G^{(i)},\mathcal{F}, n_{\text{circuit}}/2).
\end{align}

These selected circuits undergo crossover to create a new generation of offspring. We stress that, unlike previous GA-QSVM frameworks\cite{Pellow-Jarman2024}, our approach forces the offspring to be normalized under several conditions that balance the expressivity and trainability of quantum circuits. For example, (1) circuits should be shortened if their depth is longer than a pre-defined depth $d$, and (2) CX gates must be added if the number of CX gates is less than the number of required CX gates.

\begin{align}
    G^{(i+1)}\gets \text{Crossover.OnePoint}(\{G_{\text{father}}^{(i)}, G_{\text{mother}}^{(i)}\}, \text{Normalizer.Depth}(d), \ldots).
\end{align}

We use the bit-flip mutation with a mutation rate $p$. For each gate $g$, there is a $p\%$ chance that $g$ changes to a different gate belonging to the pool gates. The value $p$ is thus crucial to find an optimal circuit. If $p$ is low, GA-QSVM will get stuck at local minima; otherwise, the evolution process can be unstable. The mutated circuits also need to be normalized following the normalization condition:
\begin{align}
    G^{(i+1)}\gets \text{Mutation.BitFlip}(G^{(i+1)}, 
    \text{Normalizer.Depth}(d), \ldots).
\end{align}


\subsection{Metadata for GA-QSVM} 

In the proposed GA-QSVM framework, \emph{metadata} is defined as a tuple of hyperparameters that configure both the quantum feature map and the GA process. Formally, we denote the metadata as
\begin{align}
    \mathcal{M} = \big(\tau,n, n_{R_x}, n_{R_y}, n_{R_z}, d, n_{\text{circuits}}, n_{\text{generation}}, p_m\big),
\end{align}
where $\tau$ is the threshold, $n$ is the number of qubits, $(n_{R_x}, n_{R_y}, n_{R_z})$ represent the allocation of single-qubit rotation gates along the $x$, $y$, and $z$ axes, $d$ denotes the circuit depth, $n_{\text{circuits}}$ is the number of circuits in the population, $n_{\text{generation}}$ is the number of generations, and $p_m$ is the mutation probability. The first subset of parameters $(n, n_{R_x}, n_{R_y}, n_{R_z}, d)$ governs the expressibility of the quantum circuit, while the second subset $(n_{\text{circuits}}, n_{\text{generation}}, p_m\big)$ regulates the behavior of the genetic algorithm by controlling the balance between exploration and exploitation. By encapsulating these parameters in $\mathcal{M}$, the framework ensures reproducibility, enables systematic benchmarking across hyperparameter configurations.

\section{Numerical results}
\label{sec:evaluation}
\subsection{Dataset preparation}

We have performed QSVM with two types of quantum kernels, including PQK and FQK using Qiskit 1.3.1 \cite{javadi2024quantum} and Squlearn 0.8.4 \cite{10841386}, respectively. Regarding datasets, we used Scikit-learn 1.6.1 \cite{scikit-learn} for wine, digits, and breast-cancer datasets and Tensorflow 2.16.2 \cite{199317} for the Fashion MNIST dataset. All calculations were performed on an Intel Core i9-10920X CPU at 3.50 GHz and 120 GB of RAM. For comparison, we have used a radial basis (RBF) kernel for classical SVM, and Z/ZZ feature map with FQK and PQK for conventional QSVM. 



\begin{table}[t]
\centering
\caption{Dataset properties.}
\label{tab:dataset}
\resizebox{.65\textwidth}{!}{\begin{tabular}{|l|c|c|c|c|}
\hline
\textbf{Dataset} & \textbf{Ref} & \textbf{\#Instances} & \textbf{\#Classes} & \textbf{\#Features} \\
\hline
\textbf{Digits}        & \cite{digits} & 5620  & 10 & 64  \\
\textbf{Fashion}       & \cite{xiao2017fashion} & 70000 & 10 & 784 \\
\textbf{Wine}          & \cite{wine_109} & 178   & 3  & 13  \\
\textbf{Breast Cancer} & \cite{breast_cancer_wisconsin} & 592   & 2  & 30  \\
\hline
\end{tabular}}
\end{table}

\begin{figure*}[t]
    \centering
    \includegraphics[width=0.99\textwidth]{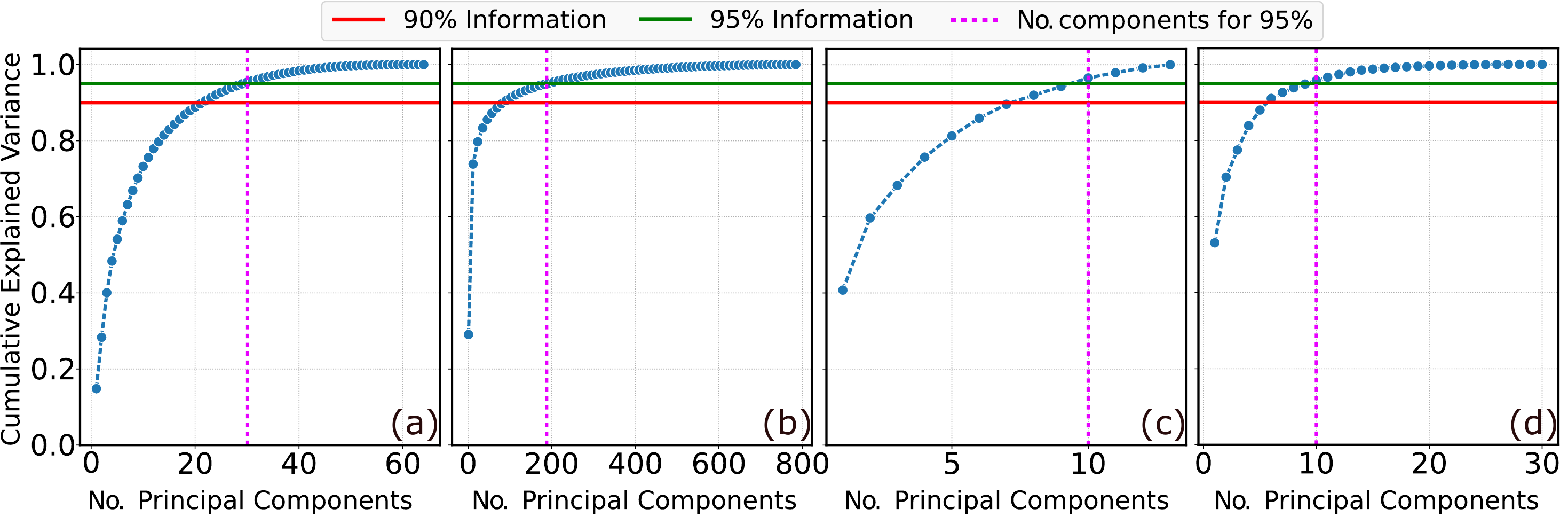}
    \caption{Cumulative explained variance of (a) \textbf{Digits}, (b) \textbf{Fashion}, (c) \textbf{Wine}, and (d) \textbf{Breast Cancer} dataset after using PCA. The number of components for 95\% cumulative explained variance for four datasets is 30, 200, 10, and 10.}
    \label{fig:pca}  
\end{figure*}

We use four datasets for the classification task, including: \textbf{Digits}, \textbf{Fashion}, \textbf{Wine}, and \textbf{Breast Cancer} as described in Table.~\ref{tab:dataset}. The maximum number of training items and testing items is $100$ and $100$ for all experiments, respectively. Only three datasets, including \textbf{Digits}, \textbf{Wine}, and \textbf{Breast Cancer} were used in GA-QSVM because of the huge number of Instances (\#Instances) from \textbf{Fashion}. Due to the limitation of the number of qubits, Principal Component Analysis (PCA) was applied to reduce the number of Features (\#Features) of data points $\{x_i\}$. Note that the number of qubits must be equal to \#Feature because we use the angle encoding method. Some first experiments are shown in Figure.~\ref{fig:pca}, the number of principal components needed to retain a significant amount of information varies greatly by dataset. The \textbf{Wine} and \textbf{Breast Cancer} datasets are less complex, requiring only $10$ components for $95\%$ information, while the \textbf{Digits} and \textbf{Fashion} datasets, especially the latter, require a much higher number of components.

\subsection{Optimal circuit searching}

\begin{figure}[t]
    \centering
    \includegraphics[width=0.6\textwidth]{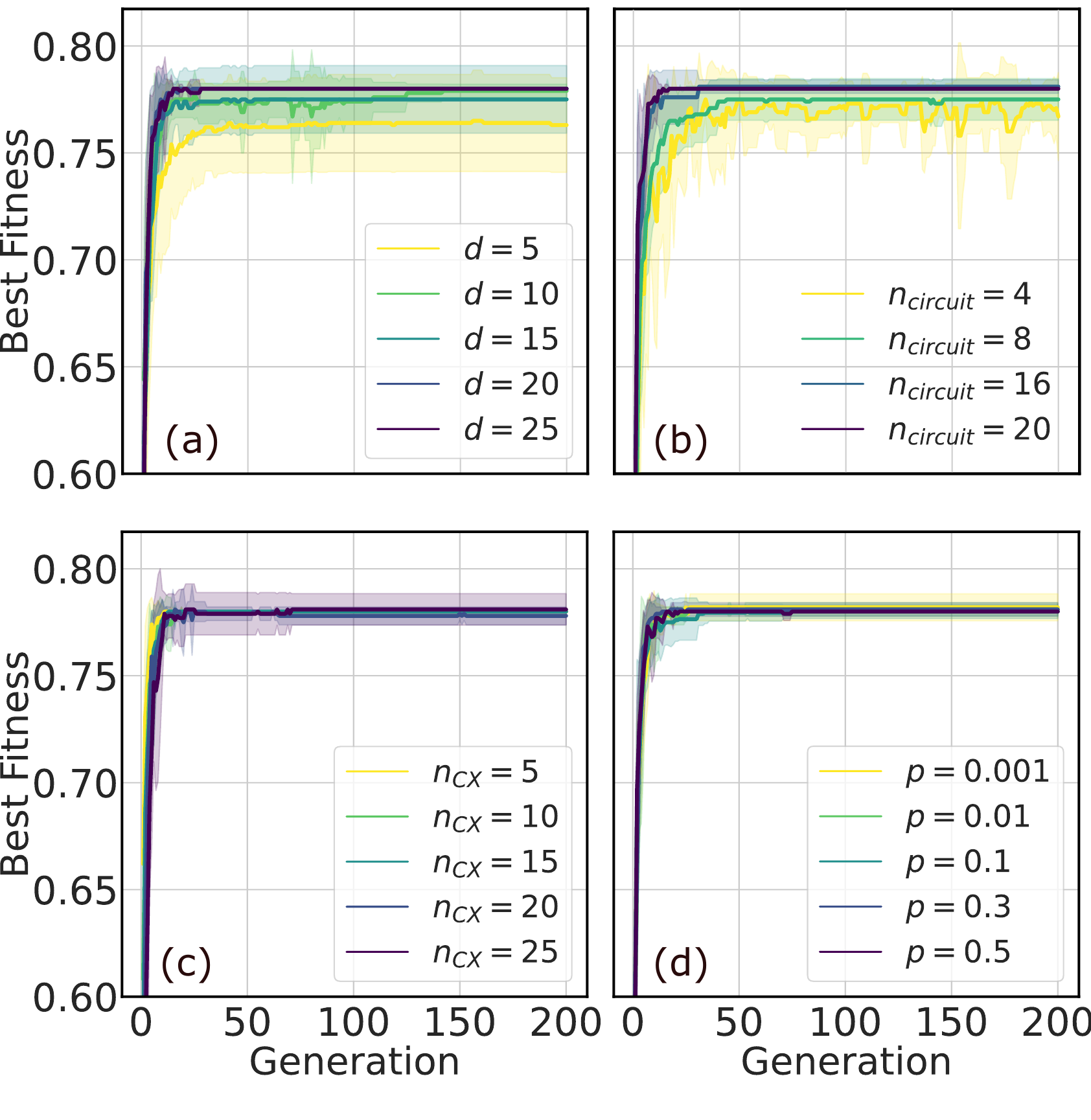}
    \caption{We conduct the survey on hyperparameters by benchmarking on \textbf{Digits} dataset, with fixed $n=5$, $n_{\text{circuit}}=16$, $p=0.1$, $d=5n$, $n_{\text{CX}}=2n$ and different configuration. (a) different $d$, (b) different $n_{\text{circuit}}$, (c) different $n_{\text{CX}}$ and (d) different $p$. Note that in GA-QSVM, we run the fitness function (QSVM) with fewer iterations to save the whole optimization time.}
    \label{fig:benchmark_hyperparameter}  
\end{figure}

As standard GA, we measure the best fitness versus generation. As an example, we plot the searching process for the digit dataset in  Figure.~\ref{fig:benchmark_hyperparameter}. The best fitness starts from a low value, becomes better through generations, and reaches an upper bound limit that comes from the resources we restrict and the dataset. We investigate how hyperparameters affect the best fitness. The benchmark results show the effect of different configurations: $n_{\text{CX}}\in[5,10,15,20,25]$, $p\in[0.001, 0.01, 0.1, 0.3, 0.5]$, $d\in[5, 10, 15, 20, 25]$, $n_{\text{circuit}}\in[4, 8, 16, 20]$.

As shown in Figure.~\ref{fig:benchmark_hyperparameter}a, the circuit depth $d$ has a noticeable effect on the performance; deeper circuits, such as $d=20$ and $d=25$, lead to a higher and more stable fitness. The size of the population $n_{\text{circuit}}$, shown in Figure.~\ref{fig:benchmark_hyperparameter}b, has a significant impact on performance and convergence
Larger populations, such as $n_{\text{circuit}}=16$ and $n_{\text{circuit}}=20$, converge more quickly and achieve a higher, more stable final fitness with less variance.
As we can see in Figure.~\ref{fig:benchmark_hyperparameter}c and d, the best fitness for different $n_{\text{CX}}$ and $p$ values are tightly close to each other with the peak fitness around $0.79$. Within the tested range, the number of CX gates and mutation probability do not significantly impact the final classification accuracy. Overall, as the depth and the population size increase, more resources are included in the GA process, leading to better circuit searching. 

\begin{figure}[t]
    \centering
    \includegraphics[width=0.99\textwidth]{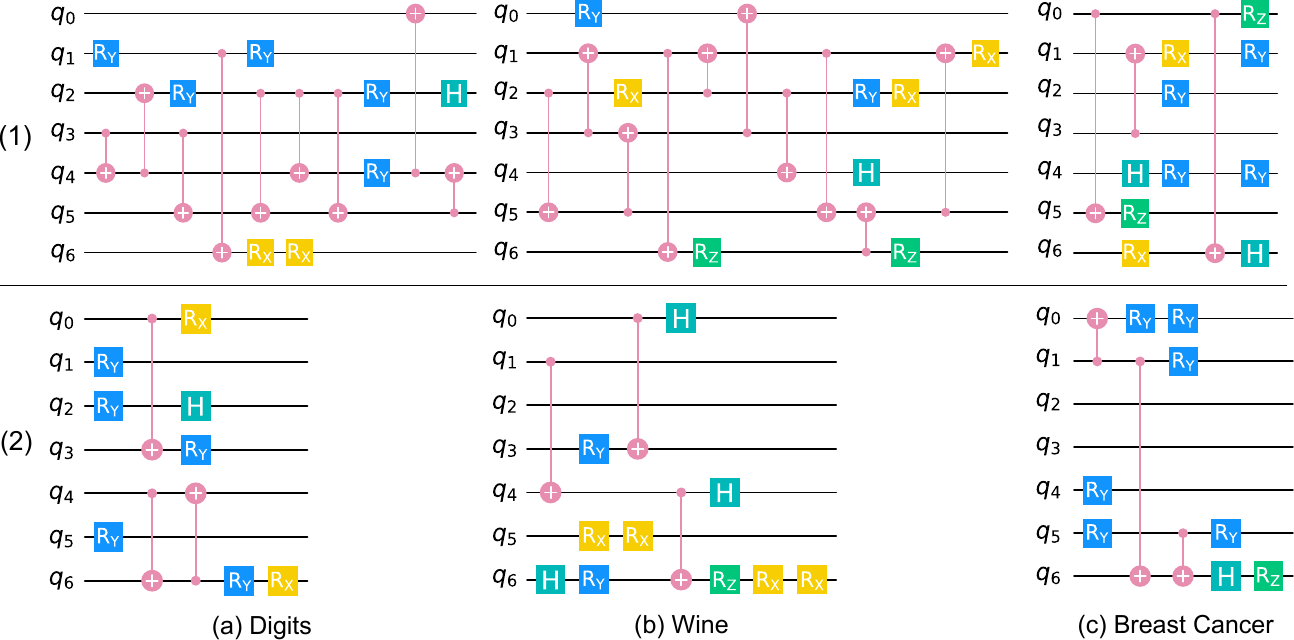}
    \caption{Optimal 7-qubit circuits from GA-QSVM for the three datasets (a) Digits, (b) Wine, and (c) Breast Cancer with two quantum kernels (1) FQK and (2) PQK.
    }
    \label{fig:ansatz}  
\end{figure}

Figure~\ref{fig:ansatz} illustrates the optimal $7$-qubit circuits selected by the GA-QSVM framework. Each subfigure (a)-(c) corresponds respectively to the Digits, Wine, and Breast Cancer datasets, while rows (1) and (2) show the best circuits produced under the FQK and PQK kernels. For all datasets, GA yields shallow circuits combining $\{R_x, R_y, R_z, H, \text{CX}\}$. Notably, the FQK-optimized circuits tend to add more CX gates, whereas PQK-optimized circuits often emphasize balance between local and non-local gates. The structural differences across datasets highlight how the GA adapts circuit topology to dataset complexity: the Breast Cancer datasets favor shallower constructions with fewer CX gates. 

\subsection{GA-QSVM performance}

\begin{figure*}[t]
    \centering
    \includegraphics[width=0.99\textwidth]{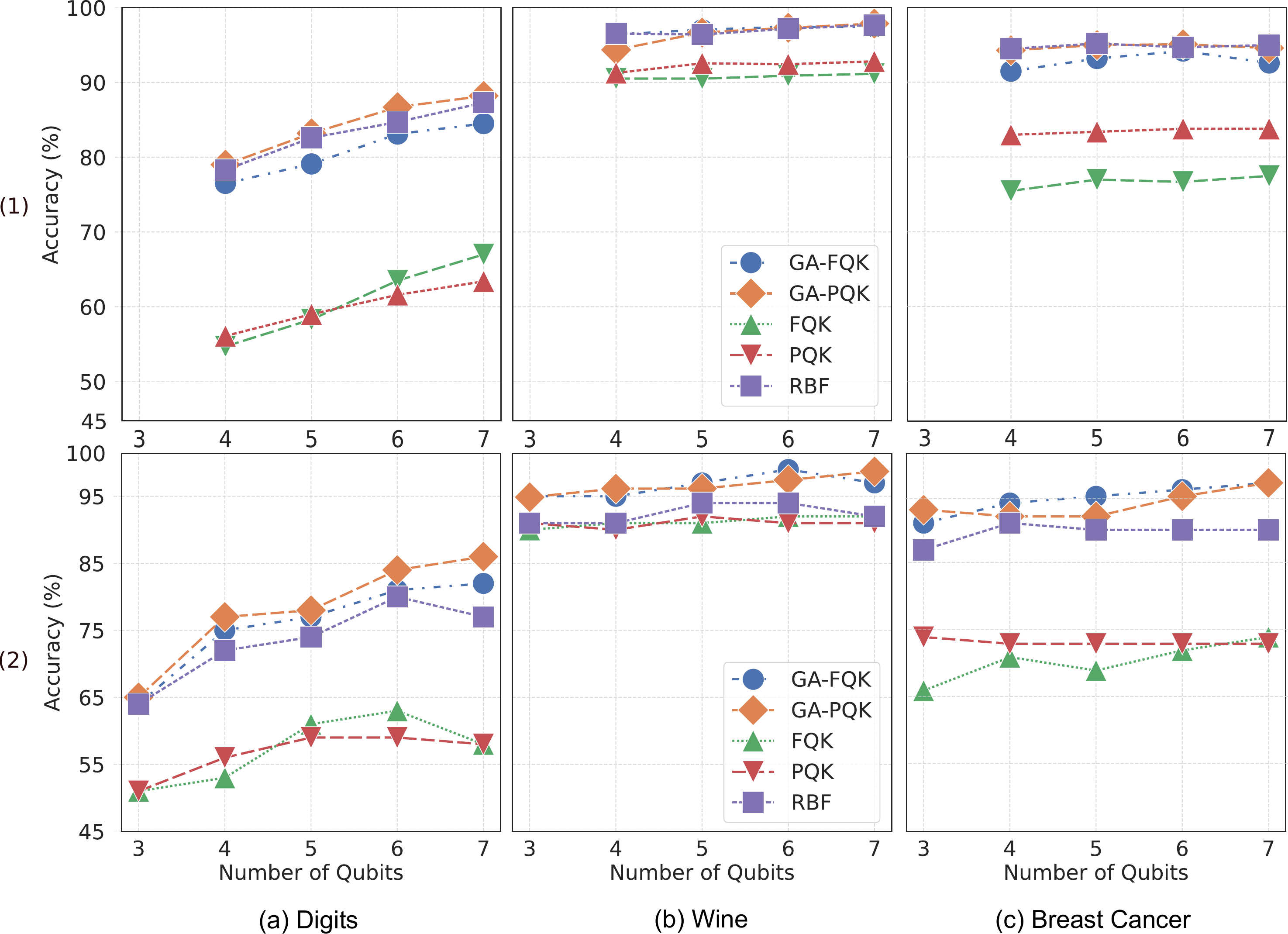}
    \caption{Accuracy between models classical SVM (RPF), existing QSVM kernel (FQK, PQK) and our optimized kernel (GA-FQK, GA-PQK) with default configuration $n_{\text{circuit}}=16$, $p= 0.1$, $d=5$, $n_{\text{CX}}=2n$; which use (1) standard train/test/val split (from $4$ to $7$ qubits) and (2) $k$-fold (from $3$ to $7$ qubits). The QSVM models with optimized kernel are run with full number of iterations.}
    \label{fig:final_res}  
\end{figure*}

Figure~\ref{fig:final_res} compares the accuracy of different methods (GA-FQK, GA-PQK, FQK, PQK, RBF) for three datasets using the standard train/test/val split (subfigure (1)) and $k$-fold cross-validation (subfigure (2)). In $k$-fold cross-validation, the QSVM is trained $k$ times through $k$ train–test splits, then the results are averaged to obtain the final accuracy \cite{10.1162/089976698300017197}. The performance of the standard train/test/val method is not stable because it depends on how the test set is split. In contrast, $k$-fold cross-validation yields a more robust estimate of the model’s generalization performance \cite{manual2013introduction}.

We can see that the conventional QSVM (FQK/PQK) with a fixed ansatz shows the worst accuracy for all cases. For the standard train/test/val method, the performance of GA-FQK/GA-PQK and RBF is comparable. However, the performance of GA-QSVM with both FQK and PQK is noticeably better than the classical kernel RBF for the $k$-fold cross-validation (subfigure (2)).
In general, the comparison highlights that our proposed methods maintain competitive or superior performance under the standard evaluation, and both quantum and classical SVMs achieve near-saturation accuracy.


To prove the expressivity of GA-QSVM, we run GA-QSVM on a base dataset, then use the solution (the best circuit) as an ansatz for QSVM on other datasets. As an example, we choose \textbf{Digits} as the source for transfer because it is a complex dataset, forcing the GA-QSVM to learn a highly expressive ansatz. This expressive structure is then expected to transfer to other datasets. Transfer learning aims to reduce the number of trials for GA-QSVM on other datasets. 

\begin{figure}[t]
    \centering
    \includegraphics[width=0.7\textwidth]{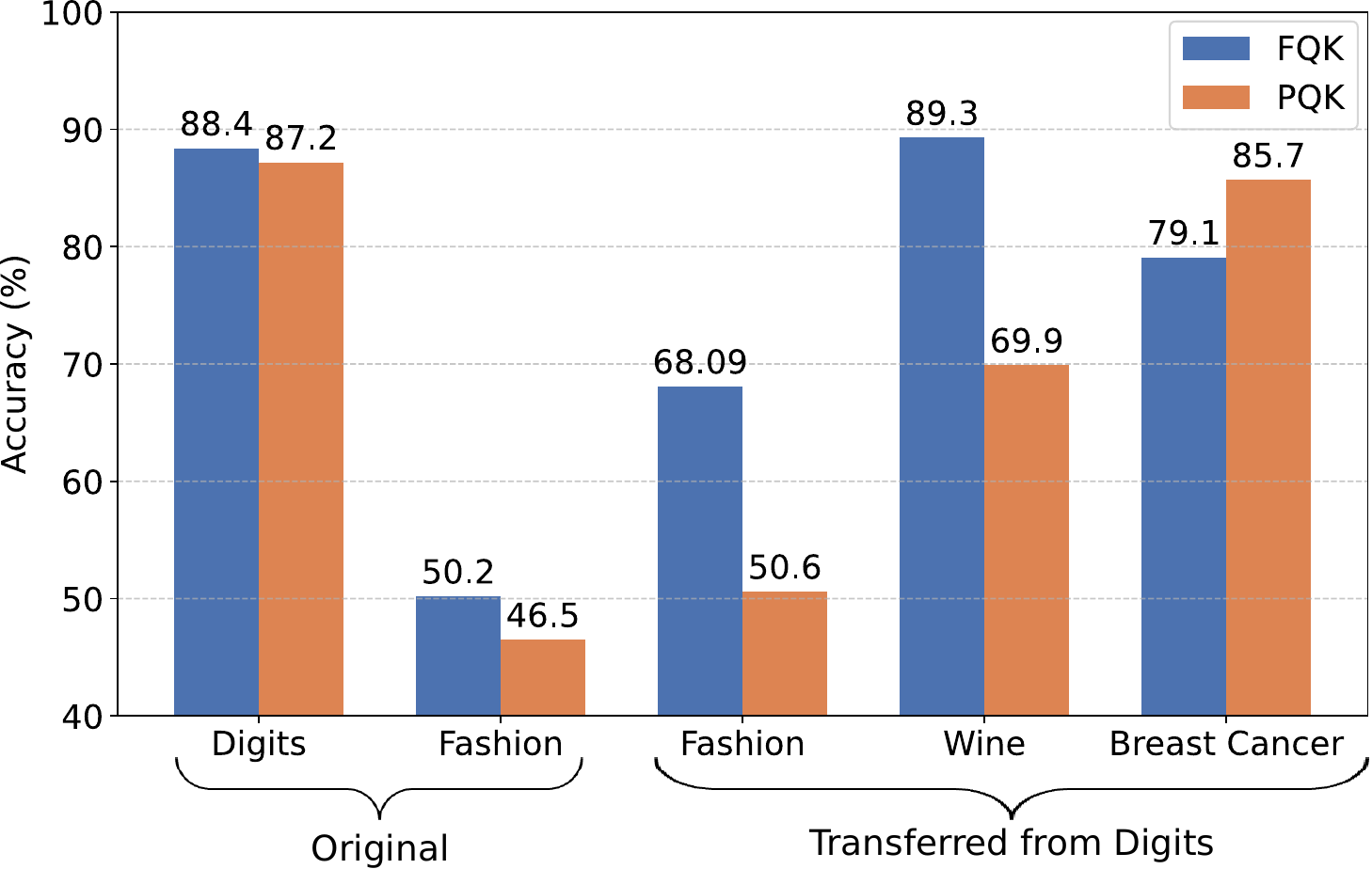}
    \caption{Accuracies from our optimized kernels on different datasets using transfer learning (right columns). The number of features (number of qubits) used in this experiment is 10.}
    \label{fig:transfer}
\end{figure}

Figure~\ref{fig:transfer} summarizes the performance of transfer learning across different target datasets when using \textbf{Digits} as the base dataset. To enhance the expressivity of the transferred circuit, we have used 10 features corresponding to 10 qubits. The first two columns are results from the GA-QSVM original training for \textbf{Digits} and \textbf{Fashion}, whereas the last three columns represent the accuracy of QSVM for \textbf{Fashion}, \textbf{Wine}, and \textbf{Breast Cancer} using the ansatz transferred from \textbf{Digits}. 

\textbf{Fashion} is a complex dataset that the classical SVM can only achieve an accuracy of 65.7\%, and QSVM/PQK with the standard Z-Feature Map (with $n=10$) has a very low accuracy of 22.9\%. Also, we can see that the \textbf{Fashion} dataset is so complex that the original GA-QSVM cannot achieve high accuracies (only 50.2\% and 46.5\% for FQK and PQK, respectively). However, when the ansatz transferred from \textbf{Digits} is used for \textbf{Fashion}, the accuracy is significantly improved. Especially, the FQK accuracy is up to 68.1\%, higher than the classical SVM (65.7\%). This improvement may arise from the ansatz learned on \textbf{Digits}, capturing richer feature correlations and offering a more favorable inductive bias than training on \textbf{Fashion} alone. For \textbf{Wine} and \textbf{Breast Cancer}, the transfer learning can also achieve reasonable accuracies. The comparison also reveals that FQK overall outperforms PQK in transfer learning, except for the \textbf{Breast Cancer} where PQK shows superior performance ($85.7$ vs $79.1$).  In general, using transfer learning can help to reduce the computational demands of GA training. Still, an open question is how to choose a proper base dataset to get an expressive ansatz for various datasets.  


\section{Conclusion}\label{sec:conclusion}

By allowing flexible metadata configuration and adaptive search, GA-QSVM overcomes the limitations of manually designed circuits and improves kernel expressivity. Numerical results on various datasets demonstrate that GA-QSVM achieves an accuracy superior to standard QSVM with a fixed quantum feature map, especially for complex data. Moreover, the transfer-learning evaluation highlights that a GA-QSVM circuit trained on one dataset can be effectively generalized across domains, thereby reducing training time for different datasets.


The GA-QSVM approach also presents limitations. The optimization process is still computationally expensive due to repeated kernel evaluations, and the stochastic nature of GA can lead to instabilities across runs. Additionally, as the number of qubits increases, circuit search and evaluation become much more difficult, restricting the scalability on near-term devices. Future work will focus on extending GA-QSVM to multi-objective optimization, enabling simultaneous control over accuracy, circuit depth, and entanglement cost. These directions may further enhance the practicality of evolutionary quantum kernel design for real-world machine learning tasks.

\section*{Acknowledgment}

This research is funded by the National Foundation for Science and Technology Development (NAFOSTED) Grant Number 103.01-2024.06. L.B.H. is supported by JSPS KAKENHI Grant Number 23K13025 and the Tohoku Initiative for Fostering Global Researchers for Interdisciplinary Sciences (TI-FRIS) of MEXT's Strategic Professional Development Program for Young Researchers.

\section*{Data and code availability}

Data are available from the corresponding authors upon reasonable request. The code is available at \url{https://github.com/vutuanhai237/GA-QSVM}.

\bibliographystyle{IEEEtran}
\bibliography{references.bib}

\end{document}